\documentclass{icrc2009}

\usepackage{graphicx}   
\usepackage[caption=false]{caption}    
\usepackage[font=footnotesize]{subfig} 
\usepackage{fixltx2e}
\usepackage{url}

\newcommand{\shorttitle}[1]%
{\markboth{Proceedings of the 31\MakeLowercase{$^{st}$} ICRC, {\L}\'{o}d\'{z} 2009}{#1} }
\newcommand{\etal}{\MakeLowercase{\textit{et al. }}} 

\usepackage{multirow}
\usepackage{amssymb}

\begin{document}
\title{Searches for WIMP Dark Matter from the Sun with AMANDA}

\author{\IEEEauthorblockN{James Braun\IEEEauthorrefmark{1} and
                           Daan Hubert\IEEEauthorrefmark{2} for the
                           IceCube Collaboration\IEEEauthorrefmark{3}}
                            \\
\IEEEauthorblockA{\IEEEauthorrefmark{1}Dept. of Physics, University of Wisconsin, Madison, WI 53706, USA}
\IEEEauthorblockA{\IEEEauthorrefmark{2}Vrije Universiteit Brussel, Dienst ELEM, B-1050 Brussels, Belgium}
\IEEEauthorblockA{\IEEEauthorrefmark{3}see http://www.icecube.wisc.edu/collaboration/authorlists/2009/4.html.}}

\shorttitle{Braun \etal AMANDA Dark Matter Searches}
\maketitle

\begin{abstract}
A well-known potential dark matter signature is emission of GeV - TeV neutrinos from annihilation of neutralinos gravitationally bound to massive objects. We present results from recent searches for high energy neutrino emission from the Sun with AMANDA, in all cases revealing no significant excess. We show limits on both neutralino-induced muon flux from the Sun and neutralino-nucleon cross section, comparing them with recent IceCube results. Particularly, our limits on spin-dependent cross section are much better than those obtained in direct detection experiments, allowing AMANDA and other neutrino telescopes to search a complementary portion of MSSM parameter space.
\end{abstract}

 
\section{Introduction}

Weakly interacting massive particles (WIMPs) with electroweak scale masses are currently a favored explanation of the missing mass in the universe.
Such particles must either be stable or have a lifetime comparable to the age of the universe, and they would interact with
baryonic matter gravitationally and through weak interactions.  The minimal supersymmetric standard model (MSSM) provides a natural
candidate, the lightest neutralino \cite{cand}.  A large range of potential neutralino masses exists, with a lower bound on the mass of the
lightest neutralino of 47 GeV imposed by accelerator-based analyses \cite{pdg}, while predictions based on the inferred dark matter
density suggest masses up to several TeV \cite{gilmore}.

Searches for neutralino dark matter include {\it direct} searches for nuclear recoils from weak interaction of neutralinos
with matter \cite{cdms, xenon} and {\it indirect} searches for standard model particles produced by neutralino annihilation.
Particularly, a fraction of neutralinos interacting with massive objects would become gravitationally bound
and accumulate in the center.  If neutralinos comprise dark matter, enough should accumulate and annihilate to produce an
observable neutrino flux.
Searches for a high energy neutrino beam from the center of the Earth \cite{wimpearth} and the Sun \cite{ic22, baksan, macro, superk, amanda}
have yielded negative results. Observations of a cosmic ray electron-positron excess by ATIC \cite{atic}, PPB-BETS \cite{ppbbets},
Fermi \cite{ferminew}, and HESS \cite{hessnew}, along with the anomalous cosmic ray positron fraction reported by PAMELA \cite{pamela},
could be interpreted as an indirect signal of dark matter annihilation in our galaxy \cite{bergstrom}.

Here we present searches for a flux of GeV--TeV neutrinos from the Sun using AMANDA.  We improve on the sensitivity of the previous
AMANDA analysis \cite{amanda} significantly and extend the latest results from IceCube \cite{ic22} to lower neutralino masses.  We
observe no neutralino annihilation signal and report limits on the neutrino-induced muon flux from the Sun and the resulting limits
on neutralino-proton spin-dependent cross section.

\section{Neutrino Detection with AMANDA}

The detection of neutrino fluxes above $\sim 50$ GeV is a major goal of the Antarctic Muon And Neutrino
Detector Array (AMANDA).  AMANDA consists of 677 optical modules embedded
1500 m to 2000 m deep in the ice sheet at the South Pole, arranged in 19 vertical strings and
occupying a volume of $\sim 0.02$ km$^3$. Each module contains a 20 cm diameter photomultiplier tube
(PMT) optically coupled to an outer glass pressure sphere.  PMT pulses (``hits") from incident Cherenkov light
are propagated to surface electronics and are recorded as an event when 6--7 hits on any one
string or 24 total hits occur within 2.5 $\mu$s.  The vast majority of the O($10^9$) events recorded
each year are downgoing muons produced by cosmic ray air showers in the
atmosphere above the South Pole.  Relativistic charged leptons produced near the detector via charged-current
neutrino interactions similarly trigger the detector, with several thousand atmospheric
neutrino induced muon events recorded per year.  The hit leading edge times, along with the known AMANDA
geometry and ice properties \cite{icepaper}, allow reconstruction of muon tracks with median accuracy
1.5$^{\circ}$ -- 2.5$^{\circ}$, dependent on zenith angle.

AMANDA operated in standalone from 2000--2006 and is currently a subdetector of the much larger ($\sim$ km$^3$)
IceCube Neutrino Observatory \cite{icecube}, scheduled for completion in 2011.  The optical module density of
AMANDA is much higher than that of IceCube, making AMANDA more efficient for low-energy muons ($\lesssim 300$ GeV)
which emit less Cherenkov light.

\section{Data Selection and Methods}

We describe two separate searches for Solar neutralinos in this proceeding.  First, we present a search using
a large data sample from 2000--2006 prepared for a high energy extraterrestrial point source search \cite{amandaps, jimthesis}.
We also present a search using data from 2001--2003, optimized to retain low energy events \cite{daanthesis}.  Both analyses are done
in two stages; first, neutrino induced muon events are isolated from the much larger background of downgoing
muons, then a search method is used to test for an excess at the location of the Sun.

\subsection{Data Selection}

While the Sun is above the horizon, neutrino-induced muons from the Sun are masked by the much larger background
of downgoing cosmic ray muons; thus, we select data during the period when the Sun is below the horizon (Mar. 21 -- Sept. 21),
resulting in 953 days livetime from 2000--2006 and 384 days from 2001--2003.
In both analyses, neutrino events are isolated by selecting well reconstructed upgoing muon tracks.  Events are first
reconstructed with fast pattern matching algorithms, and events with zenith angles $\theta < 80^{\circ}$ ($\theta < 70^{\circ}$
for the 2001-2003 analysis)
are discarded, eliminating the vast majority of downgoing muons.  The remaining events are reconstructed with
a more computationally intensive maximum-likelihood reconstruction \cite{amandareco} accurate to
1.5$^{\circ}$ -- 2.5$^{\circ}$, and again events with $\theta < 80^{\circ}$ are discarded.

\begin{figure}[t]\begin{center}
\includegraphics[width=3.0in]{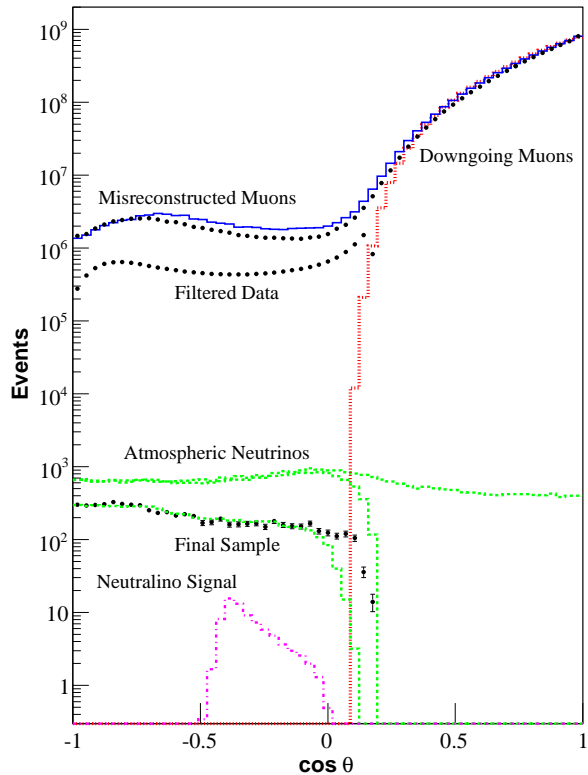}
\caption{Reconstructed zenith angles of data (circles) at trigger level, filter level ($\theta < 80^{\circ}$), and final cut level,
true (fine dotted) and reconstructed (solid) zenith angles of CORSIKA \cite{corsika} downgoing muon simulation at trigger level,
reconstructed zenith angles of ANIS \cite{anis} atmospheric neutrino simulation at trigger level, filter level, and final cut level
(dashed), and reconstructed zenith angles of a neutrino signal from the Sun at final cut level (dash-dotted).}
\label{zenith}
\end{center}\end{figure}

O($10^6$) misreconstructed
downgoing muon events remain per year, and these are reduced by cuts on track quality parameters such as
track angular uncertainty \cite{till}, the smoothness (evenness) of hits along the track \cite{amandareco}, and the
likelihood difference between the maximum-likelihood track and a forced downgoing likelihood fit using the zenith
distribution of downgoing muons as a prior \cite{amandareco}.  For the 2000--2006 analysis, 6595 events remain after
quality cuts, dominantly atmospheric neutrinos \cite{amandaps}, reduced to 4665 events by requiring dates when the Sun
is below the horizon.  Zenith distributions from 2000--2006 are shown in figure \ref{zenith}.

We consider neutralino masses from 100 GeV to 5 TeV and two extreme annihilation channels: $W^+W^-$ ($\tau^+\tau^-$ for 50 GeV)
and $b\bar{b}$, which produce high and low energy neutrino spectra, respectively, relative to the neutralino mass.
The fraction of signal events retained depends on the neutrino energy spectrum and varies from 17\% for a 5 TeV neutralino mass
and $W^+W^-$ channel to 1\% for 100 GeV and $b\bar{b}$ channel, relative to trigger level, in the 2000--2006 analysis.

The 2001--2003 analysis is a dedicated neutralino search, unlike the 2000-2006 analysis, and more consideration is given to
low energy events.  Twelve event observables are considered, and selection criteria based on these observables are
optimized separately for three signal classes, dependent on neutralino mass and annihilation channel, to maximize retention
of signal events.  The signal classes are shown below along with the number of events passing selection criteria.
\begin{center}{\small \begin{tabular}{cccc}
\hline\hline
&&&\\[-3.0mm]
Class & Channel & $m_{\chi}$ & Final Events\\[+0.5mm]
\hline
&&&\\[-3.0mm]
\multirow{2}{*}{\bf A} &$W^+W^-$  & 250 GeV -- 5 TeV & \multirow{2}{*}{670} \\
                       &$b\bar{b}$& 500 GeV -- 5 TeV & \\[+1mm]
\multirow{2}{*}{\bf B} &$W^+W^-$  & 100 GeV          & \multirow{2}{*}{504} \\
                       &$b\bar{b}$& 250 GeV          & \\[+1mm]
\multirow{2}{*}{\bf C} &$\tau^+\tau^-$  & 50 GeV           & \multirow{2}{*}{398} \\
                       &$b\bar{b}$& 50, 100 GeV  & \\[+0.5mm]
\hline\hline
\end{tabular}}\end{center}
The selection is more efficient than the 2000-2006 analysis, with 21\% of signal retained for 5 TeV, $W^+W^-$ channel, to 4\% for
100 GeV, $b\bar{b}$ channel.  The 2001--2003 analysis additionally considers 50 GeV neutralino masses, with a signal efficiency of 1\%--3\%.

\subsection{Search Method}

Both analyses use maximum-likelihood methods \cite{method} to search for an excess of events near the location of the Sun.
The data is modeled as a mixture of $n_s$ signal events from the Sun and background events from both atmospheric neutrinos and
misreconstructed downgoing muons.  The signal likelihood for the i$^{th}$ event is
\begin{equation}
\mathcal{S}_i = \frac{1}{2\pi\sigma_i^2}e^{-\frac{\psi_i^2}{2\sigma_i^2}},
\end{equation}
where $\psi_i$ is the space angle difference between the event and the Sun, and $\sigma_i$ is the event angular uncertainty
\cite{till}.  The background likelihood $\mathcal{B}_i$ is obtained from the zenith distribution of off-source data.
The full-data likelihood over all $N$ data events is
\begin{equation}
\mathcal{L} = P(Data|n_s) = \prod_{i=1}^N\Big(\frac{n_s}{N}\mathcal{S}_i + (1-\frac{n_s}{N})\mathcal{B}_i\Big)
\end{equation}
and is numerically maximized to find the best fit event excess $\hat{n}_s$.  The likelihood ratio
$-2\log\frac{\mathcal{L}(0)}{\mathcal{L}(\hat{n}_s)}$ is approximately $\chi^2$ distributed and provides a measure of
significance.  Event upper limits are set from this likelihood using the Feldman-Cousins unified construction \cite{fc}.

\subsection{Signal Simulation and Systematic Uncertainties}

Neutrino energy distributions at Earth from neutralino annihilation in the Sun are generated by DarkSUSY \cite{ds}.
For the 2000--2006 analysis, neutrino events are generated with ANIS \cite{anis}, with muons propagated using
MMC \cite{mmc}, then reweighted to the energy distributions described above.  For 2001--2003,
the DarkSUSY energy distributions are sampled by WimpSimp \cite{ws}, and muons are propagated with MMC.

Uncertainties in our signal simulation are dominated by uncertainties in optical module sensitivity and photon propagation
in ice.  These uncertainties are constrained by comparing the trigger rate of CORSIKA \cite{corsika} downgoing muon simulation
using various hadronic models with the observed AMANDA trigger rate.  The effect on signal prediction is measured by shifting
the simulated optical module efficiency by these constraints and is 10\% for $m_{\chi} = 5$ TeV, $W^+W^-$ channel, to 21\% for
$m_{\chi} = 100$ GeV, $b\bar{b}$ channel.  Other sources of uncertainty include event selection (4\%--8\%) and uncertainty
in neutrino mixing angles (5\%). For the 2000--2006 analysis, uncertainties total 13\%--24\% and are included in the limit calculation using
the method of Conrad {\it et al.} \cite{conrad} as modified by Hill \cite{hill}.  Uncertainties for 2001--2003 total 23\%--38\%
and are included in the limits assuming the worst case.

\section{Results}

\begin{figure*}[t]\begin{center}
\includegraphics[width=6in]{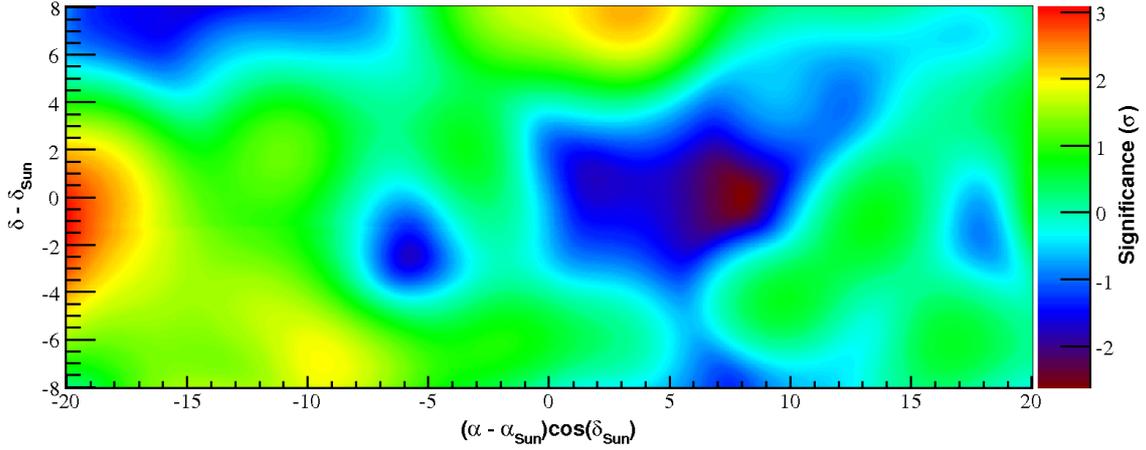}
\caption{Sun-centered skymap of event excesses from the 2000--2006 analysis.}
\label{skymap}
\end{center}\end{figure*}

The search methods are applied to the final data, and both analyses reveal no significant excess of neutrino-induced muons
from the direction of the Sun.  A Sun-centered significance skymap from the 2000--2006 analysis (figure \ref{skymap}) shows a
$0.8\sigma$ deficit from the direction of the Sun.  For the 2001--2003 analysis, a deficit of events is observed in classes A and C,
and a small excess is seen in class B.  Each excess or deficit is within the 1$\sigma$ range of background fluctuations.

\begin{figure*}[t]
\centerline{\subfloat{\includegraphics[width=3.2in]{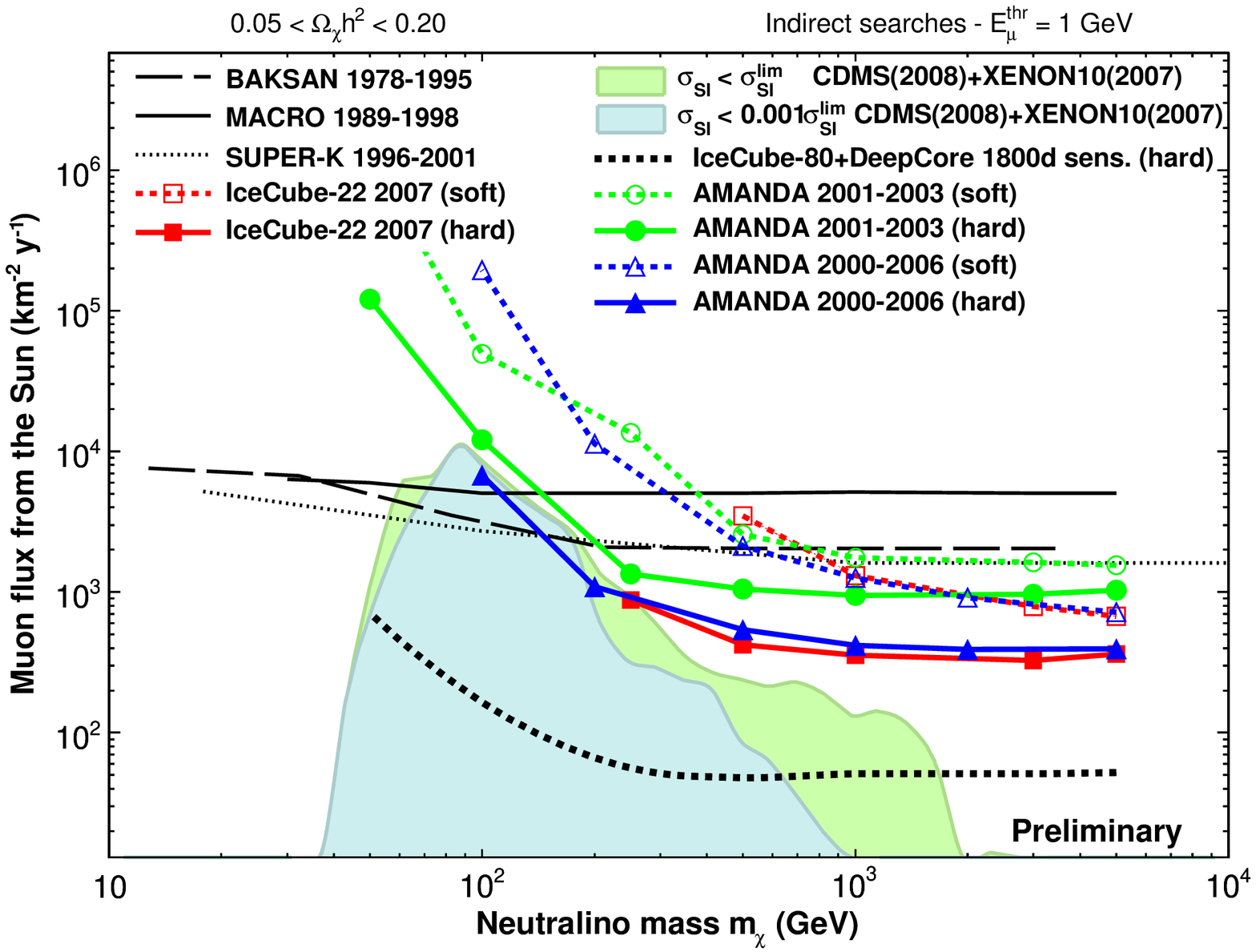} \label{sub_fig1}}
              \hfil
            \subfloat{\includegraphics[width=3.2in]{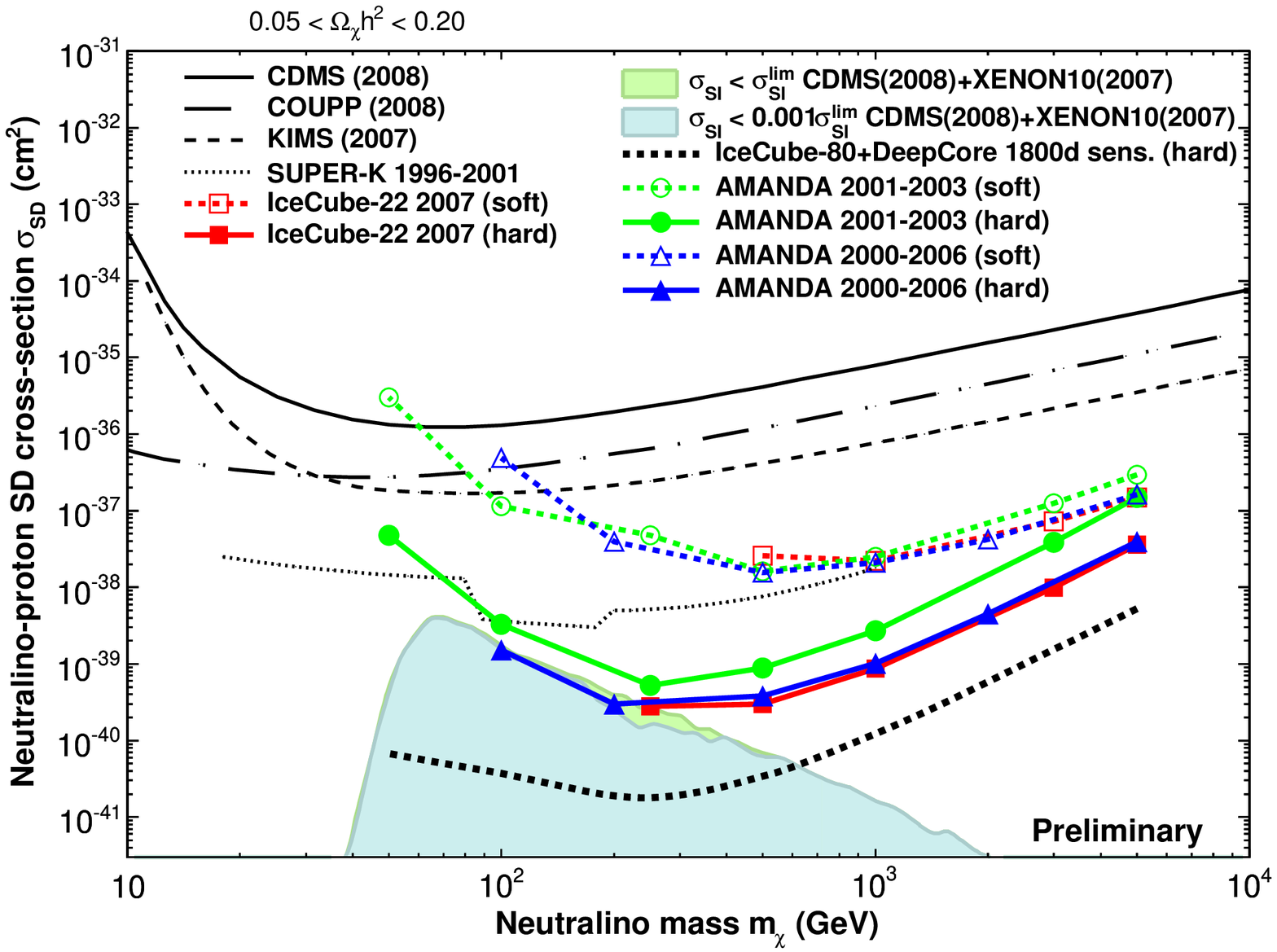} \label{sub_fig2}}
}
\caption{Preliminary limits on neutrino-induced muon flux from the Sun (left) along with limits from IceCube \cite{ic22}, BAKSAN \cite{baksan}, MACRO \cite{macro}, and
Super-K \cite{superk}, and limits on spin-dependent neutralino-proton cross section (right) along with limits from CDMS \cite{cdms}, IceCube \cite{ic22},
Super-K \cite{superk}, KIMS \cite{kims}, and COUPP \cite{coupp}.  The green shaded area represents models from a scan of MSSM parameter space not
excluded by the spin-independent cross section limits of CDMS \cite{cdms} and XENON \cite{xenon}, and the blue shaded area represents allowed models if
spin-independent limits are tightened by a factor of 1000.  The projected sensitivity of 10 years operation
of IceCube with DeepCore is shown in both figures.}
\label{limit_fig}
\end{figure*}

Upper limits on the neutralino annihilation rate in the Sun are calculated from the event upper limit $\mu_{90}$ by
\begin{equation}
\Gamma_{A} = \frac{4\pi R^2 \mu_{90}}{N_{A}\rho T_{L}V_{eff}}\Big[\int_0^{m_{\chi}} \sigma_{\nu N}\frac{dN_{\nu}}{dE}dE\Big]^{-1},
\end{equation}
where $R$ is the Earth-Sun radius, $N_{A}$ is the Avogadro constant, $\rho$ is the density of the detector medium, $T_{L}$ is the
livetime, and $\sigma_{\nu N}$ is the neutrino-nucleon cross section.  The muon neutrino energy spectrum $\frac{dN_{\nu}}{dE}$
for a given annihilation channel is obtained from DarkSUSY and includes absorption and oscillation effects from transit through the
Sun and to Earth.  The energy-averaged effective volume $V_{eff}$ is obtained from simulation.  Limits on muon flux are given by
\begin{equation}
\Phi_{\mu} = \frac{\Gamma_{A}}{4\pi R^2}\int_{1~GeV}^{m_{\chi}}\frac{dN_{\mu}}{dE}dE,
\end{equation}
and limits on neutralino-proton cross section are calculated according to \cite{cs}.  These quantities are tabulated in
table \ref{limit_table} for the more restrictive of the two analyses.  Muon flux limits, assuming a 1 GeV threshold on muon energy,
and spin-dependent cross section limits are shown in figure \ref{limit_fig} for both analyses.

\begin{table*}[t]\begin{center}
\begin{tabular}{cc|cc|c|c|cc}
\hline\hline
&&&&&&\\[-2.5mm]
$m_{\chi} (GeV)$& Channel & $V_{eff} (m^3)$ & $\mu_{90}$ & $\Gamma_{A} (s^{-1})$ & $\Phi_{\mu} (km^{-2} y^{-1})$ & $\sigma^{SI} (cm^2)$ & $\sigma^{SD} (cm^2)$ \\[+0.5mm]
\hline
&&&&&&\\[-2.5mm]
\multirow{2}{*}{50}  & $\tau^+\tau^-$ & $4.31\times 10^3$ & 6.2 & $2.11\times 10^{25}$ & $1.21\times 10^5$ & $1.84\times 10^{-40}$ & $4.80\times 10^{-38}$ \\
&$b\bar{b}$ & $8.62\times 10^2$ & 8.4 & $1.32\times 10^{27}$ & $1.32\times 10^6$ & $1.15\times 10^{-38}$ & $3.01\times 10^{-36}$ \\ \hline &&&&&&\\[-2.5mm]
\multirow{2}{*}{100} & $W^+W^-$ & $2.87\times 10^4$ & 4.5 & $1.88\times 10^{23}$ & $6.75\times 10^3$ & $3.40\times 10^{-42}$ & $1.52\times 10^{-39}$ \\[+0.5mm]
&$b\bar{b}$ & $8.65\times 10^3$ & 4.5 & $1.42\times 10^{25}$ & $4.94\times 10^4$ & $2.56\times 10^{-40}$ & $1.14\times 10^{-37}$ \\ \hline &&&&&&\\[-2.5mm]
\multirow{2}{*}{200} & $W^+W^-$ & $3.42\times 10^5$ & 4.0 & $9.81\times 10^{21}$ & $1.09\times 10^3$ & $4.23\times 10^{-43}$ & $2.98\times 10^{-40}$ \\[+0.5mm]
&$b\bar{b}$ & $9.80\times 10^3$ & 4.5 & $1.29\times 10^{24}$ & $1.13\times 10^4$ & $5.56\times 10^{-41}$ & $3.92\times 10^{-38}$ \\ \hline &&&&&&\\[-2.5mm]
\multirow{2}{*}{500} & $W^+W^-$ & $1.31\times 10^6$ & 3.7 & $2.07\times 10^{21}$ & $5.39\times 10^2$ & $3.51\times 10^{-43}$ & $3.81\times 10^{-40}$ \\[+0.5mm]
&$b\bar{b}$ & $8.87\times 10^4$ & 4.0 & $8.52\times 10^{22}$ & $2.12\times 10^3$ & $1.45\times 10^{-41}$ & $1.57\times 10^{-38}$ \\ \hline &&&&&&\\[-2.5mm]
\multirow{2}{*}{1000} & $W^+W^-$ & $2.18\times 10^6$ & 3.6 & $1.39\times 10^{21}$ & $4.18\times 10^2$ & $7.82\times 10^{-43}$ & $1.01\times 10^{-39}$ \\[+0.5mm]
&$b\bar{b}$ & $2.14\times 10^5$ & 4.0 & $2.89\times 10^{22}$ & $1.26\times 10^3$ & $1.63\times 10^{-41}$ & $2.10\times 10^{-38}$ \\ \hline &&&&&&\\[-2.5mm]
\multirow{2}{*}{2000} & $W^+W^-$ & $2.38\times 10^6$ & 3.6 & $1.56\times 10^{21}$ & $3.90\times 10^2$ & $3.19\times 10^{-42}$ & $4.52\times 10^{-39}$ \\[+0.5mm]
&$b\bar{b}$ & $3.53\times 10^5$ & 3.9 & $1.46\times 10^{22}$ & $9.10\times 10^2$ & $2.98\times 10^{-41}$ & $4.23\times 10^{-38}$ \\ \hline &&&&&&\\[-2.5mm]
\multirow{2}{*}{5000} & $W^+W^-$ & $2.07\times 10^6$ & 3.6 & $2.20\times 10^{21}$ & $3.94\times 10^2$ & $2.66\times 10^{-41}$ & $3.97\times 10^{-38}$ \\[+0.5mm]
&$b\bar{b}$ & $4.59\times 10^5$ & 3.7 & $8.91\times 10^{21}$ & $7.17\times 10^2$ & $1.08\times 10^{-40}$ & $1.61\times 10^{-37}$ \\[+0.5mm]
\hline\hline
\end{tabular}
\caption{Effective volume,  event upper limit, and preliminary limits on neutralino annihilation rate in the Sun, neutrino-induced muon flux from the Sun,
and spin-independent and spin-dependent neutralino-proton cross section for a range of neutralino masses, including systematics.}
\label{limit_table}
\end{center}\end{table*}

\section{Discussion}

These limits extend the latest IceCube limits to lower neutralino masses and are now beginning to exclude neutralino spin-dependent
cross sections allowed by direct detection experiments (figure \ref{limit_fig}).  A 1000-fold improvement over current direct-detection limits
\cite{cdms, xenon} does not significantly constrain allowed spin-dependent cross sections; thus, neutrino telescopes will continue to observe
a complementary portion of MSSM parameter space over the next several years.
IceCube is currently operating with 59 strings and will contain 86 strings when complete in 2011.  The DeepCore extension to IceCube \cite{deepcore},
six strings with tighter string spacing (72 m), tighter optical module spacing (7 m), and higher PMT quantum efficiency, will be complete in 2010.
DeepCore will significantly enhance the sensitivity of IceCube to low energy muons, extending the reach of IceCube to lower neutralino masses.

\end{document}